# Efficient Resource Allocation in Resource provisioning policies over Resource Cloud Communication Paradigm

Gaurav Raj[1], Ankit Nischal[2]

[1]Asst. Prof., Lovely Professional University,
er.gaurav.raj@gmail.com
[2]M. Tech. Student, Lovely Professional University, Phagwara
nischal.ankit@hotmail.com

*Abstract*.

*Optimal resource utilization for executing tasks within the cloud is one of the biggest challenges. In executing the task over a cloud, the resource provisioner is responsible for providing the resources to create virtual machines. To utilize the resources optimally, the resource provisioner has to take care of the process of allocating resources to Virtual Machine Manager (VMM). In this paper, an efficient way to utilize the resources, within the cloud, to create virtual machines has been proposed considering optimum cost based on performance factor. This performance factor depends upon the overall cost of the resource, communication channel cost, reliability and popularity factor. We have proposed a framework for communication between resource owner and cloud using Resource Cloud Communication Paradigm (RCCP). We extend the CloudSim[2] adding provisioner policies and Efficient Resource Allocation (ERA) algorithm in VMM allocation policy as a decision support for resource provisioner.*

## 1 Introduction

It's been some time that the term cloud computing is being in the existence. But still it emerges as a new technology as lot of research is being progressed in this field. Cloud Computing provides an illusion of infinite available resources [4]. These resources are available with the resource owner (Res_O). These resources include Processing Elements (PEs), RAM, memory and bandwidth (BW).

### 1.1 Execution of the task

   i.    User assigns the task to be executed over cloud.
   ii.   The task is received by cloud coordinator (CC).
   iii.  Cloud coordinator forwards the task over datacenters (DC).

  iv. Datacenters contains number of unfixed hosts consisting of pool of virtual machines (VM).
  v. These hosts can be configured or deleted as per the demand.

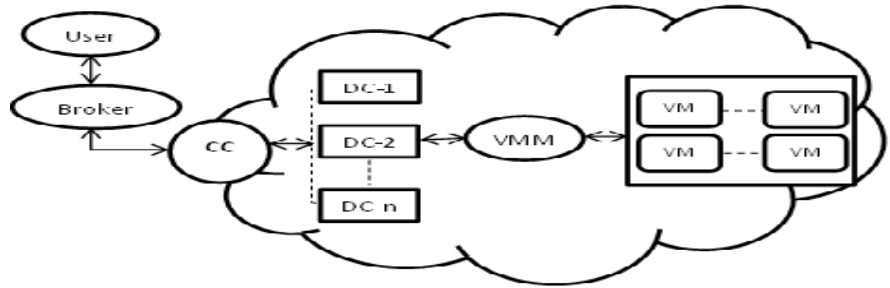

Fig. 1. Task execution over cloud

## 1.2 Allocation of resources for execution of task:

  i. VMM asks for resources from the resource provisioner by sending the task requirements..
  ii. Resource provisioner checks the availability of resources with Resource Owner.
  iii. If the resources are available, the resource owner grants the access permission to use the resources to resource provisioner.
  iv. Resource provisioner further provides access of the resources for creation of virtual machines.

## 2  Related Work

**Gaurav Raj [1]** proposed an Efficient *Broker Cloud Management* to find a communication link with minimum cost between broker and cloud. Author proposed a new broker cloud communication paradigm, explaining communication mechanism between broker and cloud using cloud exchange. Optimum route was obtained on the basis of cost factor considering hops count, bandwidth and network delay using *Optimum Route Cost Finder algorithm*.
**Rodrigo N. Calheiros et al. [2]** proposed *CloudSim simulator framework* which helped in analyzing results over large scale cloud computing infrastructure. Different allocation policies are discussed with wide scope for researchers to extend them and analyze the results. CloudSim helps in simulating the overall working contained over cloud for executing the task. It also provides extending features such as resource scheduling, cloudlet scheduling, load balancing etc…
**Shuai Zhang et al. [3]** discussed the future of cloud computing, the rise and revolution that this technology is capable of bringing in the near future. Virtualization is considered as one main characteristic of cloud computing, with which hardware, software or resources available in the cloud can be virtualized. Authors aimed cloud computing to transform the current IT industry into cloud based infrastructure and services.

## 3 Proposed Work

Most of the time, the VMMs are not optimally assigned the resources, which leads to improper utilization of resources. Cost of executing the whole task also increases due to increase in execution time and costly resource allocation in place of more efficient resources, which can be used according to the task. Load on datacenters increase because of unsorted distribution of resources while creating virtual machines.

### 3.1 Resource Cloud Communication Paradigm

*Fig. 2.* shows how the resources in RCCP are allocated to VMM.

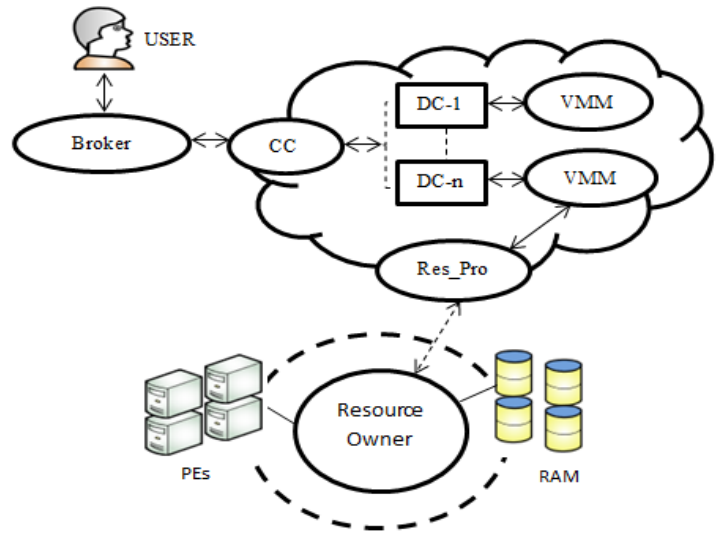

Fig. 2. Resource Cloud Communication Paradigm

**STEPS in allocating resources within RCCP**

I. Res Owners provide details of available resources to *Resource Provisioner (Res_Pro)*.
II. Res_Pro creates a **Resource Allocation List (RAL)** containing details of available resources of Res Owners.
III. VMM asks for resources from the Res_Pro by sending the requirements.
IV. Res_Pro checks the RAL.
V. If resources are available in RAL, then Res_Pro provides access to resources.
VI. If resources are not available in RAL, then Res_Pro checks availability of resources with Res Owners.
VII. Res Owners, in return, sends resource availability acknowledgements to Res_Pro.
VIII. Res Owner selects the resources considering the cost factor.
IX. Res_Pro requests for the selected resources from Res Owner.
X. Res Owner provides access to resources to Res_Pro.
XI. Res_Pro updates the RAL after getting access to resources.
XII. Res_Pro then provide access to VMM.

International Journal on Cloud Computing: Services and Architecture(IJCCSA),Vol.2, No.3, June 2012

*Sequence Diagram* illustrating resource allocation from Res Owners to VMMs

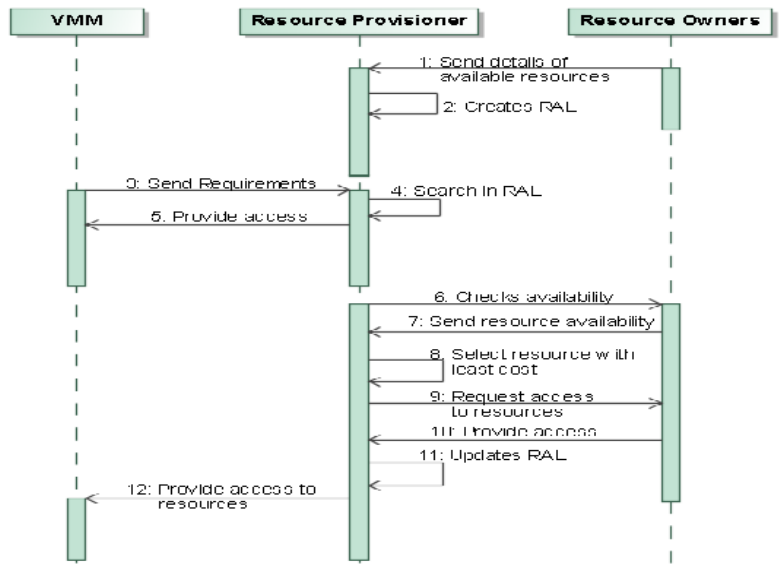

Fig. 3.  Sequence diagram for Resource Allocation Process

Resource Provisioner keeps *Optimality check* which includes certain factors (as shown in fig. 4.).

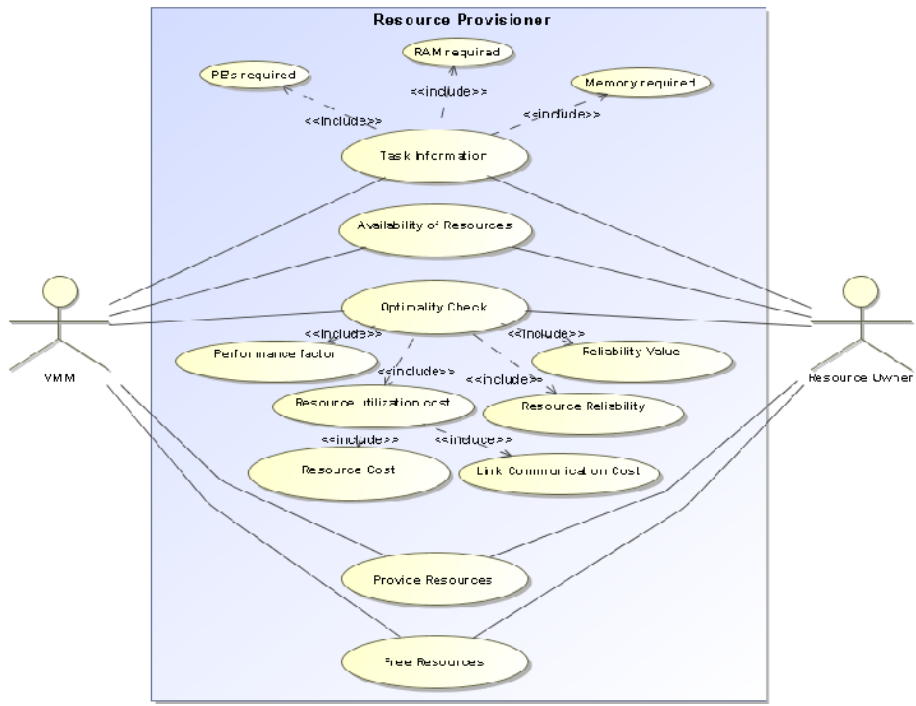

Fig. 4. Use Case Diagram of Resource Provisioner implementing factors

We have proposed an *Efficientresource allocation mechanism (ERAM)* in which *Efficient resource allocation (ERA)algorithm* is used for allocating resources. This algorithm considers certain factors while allocating the resources and hence results in efficient resource utilization. These factors are:

I. **Performance factor (PF)**
II. **Resource utilization cost ($R_{UC}$)**

    a) **Resource Cost ($R_C$)**
    b) **Link Communication Cost ($L_{CC}$)**

III. **Resource Reliability (R)**
IV. **Popularity Value (PV)**

**Performance factor (PF):** The past performance of resources within the resource pool is a crucial factor for allocating correct resources to VMM. The overall execution of the task considering PF reduces execution time and saves cost as well. Performance Factor depends upon:

i. Resource Utilization Cost
ii. Reliability
iii. Execution Time

**Resource Utilization Cost (RUC):** The cost for allocating resources to the virtual machines depends upon two factors:

I. **Resource Cost (RC):** Cost of resource which is requested.
II. **Link Communication Cost (LCC):** cost in bringing the access of resources from Res Owners to VMM. It is dependent on 3 factors [2]

    a. **Hops Count---**Response from the path, from which the resources will pass through.
    b. **Bandwidth---**The bandwidth between two nodes while fetching the resources from resource pool.
    c. **Network Delay---**Delay in sending request and receiving response.

**Resource Reliability (RR):** Reliability of any resource is measured on a reliability scale thus helping in allocating efficient resources to Virtual Machines.

**Popularity Value (PV):** Depends upon the Execution time ($E_T$). Execution time is the time taken by the resource for executing a task.
*Popularity Value is inversely proportional to Execution Time*

$$PV \; \frac{1}{E_T}$$

**Formula for calculating PF:**

PF is *directly proportional to Reliability*

$$PF \propto R \qquad (1)$$

PF is *inversely proportional to Cost (C)*

$$PF \propto \frac{1}{C} \qquad (2)$$

PF is *inversely proportional to* $E_T$

$$PF \propto \frac{1}{E_T} \qquad (3)$$

From equation (1), (2) and (3), PF can be written as:

$$PF \propto \frac{R}{C \cdot E_T}$$

$$or \;\; PF = \beta \frac{R}{C \cdot E_T}$$

where $\beta$ is Performance Factor Co − efficient

## 3.2 Efficient resource allocation (ERA) algorithm

**Case A: VM Creation**

**Step 1:** VMM analyses the task and asks for resources required to create VMs from Res_Pro.

$$VMM\_SendRequest\;[Resources] \rightarrow Res\_Pro$$

**Step 2: Res_Pro** checks **RAL**

       IF (Resources_available)

       Then Provide access to VMM and Exit

        ELSE

       Continue

**Step 3:** Res_Pro checks availability of resources with Res_O
**Step 4:** Res_O sends resource availability information to Res_Pro
**Step 5:** Res_Pro selects efficient resource based upon *cost factor (CF)*.

$$CF = R_C + L_{CC}$$

**Step 6:** Res_Pro requests Res_O for selected resources.

International Journal on Cloud Computing: Services and Architecture(IJCCSA),Vol.2, No.3, June 2012

**Step 7:** Res_O grants access permission to use the resources to Res_Pro

$$Res\_Pro\ [Resource\_Access]\quad true$$

**Step 8:** Res_Pro updates the RAL

$$RP\quad updates\ [RAL]$$

**Step 9:** Res_Pro, based upon the Performance Factor (PF), provides access to resources to VMM.

$$PF = \beta[\frac{R}{C\ E_T}]$$

**Step 10:** VMM creates VMs

$$VMM\quad creates\ [VMs]$$

**Case B: VM Deletion**
**Step 11: Check VM_Load** over VMM in VM pool

        IF VM_Load increases
        Then goto Step 1
        ELSE IF VM_Load decreases
        Then

        a) VMM searches for VMs with **least PF** and **higher $E_T$**
        b) VMM redirects the cloudlets of that VM to other VMs
        c) VMM deletes that particular VM.

**Step 12:** VM executes the cloudlets till the task is completed
**Step 13:** VMM returns the access to Res_Pro
**Step 14:** Res_pro updates RAL

## 3 Conclusion and Future work

RCCP helps in determining the overall allocation of resources with the help of Resource provisioner. ERA algorithm provides complete optimality check which results in reducing the wastage of resources. Resources available with different resource owners are selected with the help of *Performance Factor (PF)* which results in cost optimization. As PF is a function of Execution time, hence time taken to execute any task over the cloud is also reduced. As a part of future work, we will first compare the results obtained by the combination of above mentioned factors. ERA algorithm can be tested over different scenarios with varying load. More work will be done to reduce the execution time of task over VM. Introducing effective decision support mechanism in selecting available resources in ERA algorithm will make it more powerful.

**Authors**

1. Ankit Nischal

Mr. Ankit Nischal is pursuing his B.Tech-M.Tech (integrated) degree in Information Technology from Lovely Professional University, Punjab.

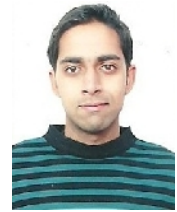

2. Gaurav Raj

Mr. Gaurav Raj is completed his Master's degree in Computer Science from Motilal Nehru National Institute of Technology, Allahabad and is currently designated as Assistant professor in Lovely Professional University, Punjab.

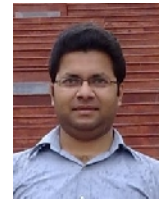